# Structure and equation of state of tetragonal boron subnitride $B_{50}N_2$

Kirill A. Cherednichenko and Vladimir L. Solozhenko*

*LSPM–CNRS, Université Paris Nord, 93430 Villetaneuse, France*

**Abstract**

*New boron subnitride $B_{50}N_2$ has been synthesized by crystallization from the B–BN melt at 5 GPa, and its structure has been refined using Rietveld analysis. $B_{50}N_2$ crystallizes in the tetragonal space group $P\bar{4}n2$ with unit cell parameters $a_0$ = 8.8181(2) Å and $c_0$ = 5.0427(10) Å. Quasi-hydrostatic compression of two boron subnitrides, $B_{50}N_2$ and $B_{13}N_2$, has been studied to 30 GPa at room temperature in a diamond-anvil cell using synchrotron X-ray diffraction. No pressure-induced phase transitions have been observed. A fit of experimental p-V data to the Murnaghan equation of state yielded $B_{50}N_2$ and $B_{13}N_2$ bulk moduli of 167(2) and 205(2) GPa, respectively, with fixed first bulk modulus pressure derivative of 4.0.*

## I. Introduction

Boron-rich compounds are an object of considerable interest due to low density, superior thermal stability, chemical resistance, promising mechanical and electronic properties that offers potential for their use as advanced engineering and smart functional materials [1-8]. However, at the present time there is a lack of experimental data on the properties of boron-rich solids, especially, under pressure.

Recent experimental study of chemical interaction in the B–BN system at high pressures and high temperatures [9] resulted in synthesis of two new boron subnitrides: $B_{13}N_2$ and "$B_{50}N_2$" (nitrogen-doped α-tetragonal boron that differs from the $B_{48}B_2N_2$ phase synthesized at ambient pressure [10,11]). $B_{13}N_2$ is low-compressible boron-rich solid [9] and is expected to exhibit hardness of about 40 GPa [12] i.e. belongs to the family of superhard phases. While structure and some properties of $B_{13}N_2$ have been already investigated [9,13,14], "$B_{50}N_2$" is not studied at all.

In the present work we performed the Rietveld refinement of the crystal structure of tetragonal $B_{50}N_2$ and studied compression behavior of two boron subnitrides, $B_{50}N_2$ and $B_{13}N_2$, up to 30 GPa at room temperature using angle-dispersive synchrotron X-ray diffraction.

## II. Experimental

Polycrystalline samples containing both boron subnitrides, $B_{13}N_2$ and $B_{50}N_2$, were synthesized in a toroid-type apparatus at ~5 GPa by quenching of B–BN melts of different compositions from 2630 K according to the procedure described elsewhere [9]. Powders of crystalline β-rhombohedral boron (99%, Alfa Aesar) and hexagonal graphite-like boron nitride (hBN) (99.8%, Johnson Matthey

* Corresponding author: vladimir.solozhenko@univ-paris13.fr

GmbH) were used as starting materials. The recovered samples were treated with 7N nitric acid (ACS, Alfa Aesar) for 20 min at 370 K in order to remove unreacted boron, washed with deionized water and dried at 400 K. X-ray powder diffraction study (Equinox 1000 Inel diffractometer, Cu*K*α1 radiation) showed that all samples contain well-crystallized $B_{50}N_2$ and $B_{13}N_2$ with traces of hBN and $B_6O$. Lattice parameters of boron subnitrides (Table I) were calculated by least-squares fits to the indexed 2θ-values using Maud software [15]; high purity silicon ($a$ = 5.431066 Å) has been used as an internal standard.

Compressibility experiments were conducted in a Le Toullec membrane diamond anvil cell [16] at room temperature. A powder sample containing 40 vol.% $B_{50}N_2$ and 60 vol.% $B_{13}N_2$ was loaded in the hole drilled in the pre-indented rhenium gasket. Neon was used as pressure transmitting medium to maintain quasi-hydrostatic conditions. The sample pressure was determined from equation of state of solid neon [17]; the maximum pressure uncertainty was estimated to be within ±1 GPa.

*In situ* X-ray diffraction measurements were carried out at "Xpress" beamline of the Elettra synchrotron (Trieste). High-brilliance synchrotron radiation from a multipole superconducting wiggler was set to a wavelength of 0.4957 Å using a channel-cut Si (111) monochromator and focused down to 20 μm. Angle-dispersive X-ray diffraction patterns were collected with on-line MAR 345 image plate detector. The exposure time was 600 seconds. The patterns were processed with FIT2D software [18]; unit cell parameters at each pressure (see Table II) have been refined by the Le Bail method using PowderCell software [19].

## III. Results and Discussion

### *A. Crystal structure of $B_{50}N_2$*

To establish the crystal structure of $B_{50}N_2$, the Rietveld refinement of powder X-ray diffraction pattern acquired at ambient conditions was performed (Fig. 1). The background was approximated by the 5-order polynomial. No satisfactory profile fit convergence was obtained in the preliminary runs with atom coordinates of the $B_{48}B_2N_2$ structure [12] taken as initial values. Further refinement was based on the structure of α-tetragonal boron [20]. It was eventually found out that $B_{50}N_2$ has a tetrahedral unit cell with $P\bar{4}n2$ space group (N 118) containing 7 independent boron atoms (in *2a* and *8i* Wyckoff positions) and 1 independent nitrogen atom (in *2b* Wyckoff position). The atomic structure information is presented in Table III. The refinement of the boron atoms coordinates did not have any significant influence on the profile fit convergence and hence the coordinates proposed for α-tetragonal boron [20] were used. The refinement of N1 coordinates showed almost no deviation from *2b* Wyckoff position (0,0,1/2). Since B2-B7 atoms constitute $B_{12}$ icosahedral clusters their total atom site occupancies were fixed to 1.0 by default. The B1 and N1 atom site occupancies were refined, however, they were also found to be close to 1 (Table III). Thus, the performed Rietveld refinement confirmed the assumed stoichiometry of tetragonal boron subnitride i.e. $B_{50}N_2$. The X-ray density of $B_{50}N_2$ (2.408 g/cm$^3$) was found to be lower than that of $B_{13}N_2$ (2.660 g/cm$^3$) (Table I).

As one can see from Fig. 1, the proposed crystal structure of $B_{50}N_2$ is in a good agreement with experimental diffraction pattern except of five very week lines (marked by red stars) that cannot be attributed to any phase of the B–N system. The final reliability factor $R_{wp}$ was converged to 10.45 which indicates the satisfactory refinement level for the compound of the 2nd period elements. The structure of $B_{50}N_2$ (different projections of the unit cell are presented in Fig. 2) is built by the $B_{12}$-icosahedra. The nitrogen atoms have a quasi-tetrahedral environment similar to interstitial atoms in other boron-rich compounds i.e. $B_{12}C_3$, $B_{12}O_2$, $B_{13}N_2$, $B_{12}P_2$, etc. Thus, new boron subnitride $B_{50}N_2$ belongs to the family of boron-rich solids.

### *B. Equation of state of $B_{50}N_2$*

The experimental $a/a_0$ and $c/c_0$ data are presented in Table II and Fig. 3. A one-dimensional analog of the Murnaghan equation of state [21] of the form

$$r = r_0\left[1 + P\left(\frac{\beta_0'}{\beta_0}\right)\right]^{-\frac{1}{\beta_0'}} \tag{1}$$

was used to approximate the nonlinear relations between the lattice parameters and pressure. Here $r$ is the lattice parameter (index 0 refers to ambient pressure); $\beta_0$ is the axial modulus and $\beta_0'$ is its first pressure derivative. The $\beta_{0,a}$ and $\beta_{0,c}$ axis moduli that best fit the experimental data are 600(38) and 416(15) GPa, respectively. The corresponding pressure derivatives $\beta_{0,a}'$ and $\beta_{0,c}'$ are 8(4) and 11(2). The axial moduli can be then easily converted in the linear compressibilities ($k_r$) by the following expression:

$$k_r = \beta_{0,r}^{-1} = \left(\frac{d \ln (r)}{d P}\right)_{P=0} \tag{2}$$

The $k_r$ values for $a$ and $c$ directions are $1.67(11)\times10^{-3}$ and $2.40(9)\times10^{-3}$, respectively, thus, $B_{50}N_2$ is more compressible along the $c$-axis.

The axial moduli and linear compressibilities of $B_{50}N_2$ have been compared with the corresponding values of $B_{13}N_2$ obtained in the present work and previous study [9] (Table I). Tetragonal $B_{50}N_2$ was found to be more compressible along the $a$ axis, while compressibilities along $c$ axis are almost the same for both subnitrides. In spite of the fact that the lattice parameters of $B_{13}N_2$ synthesized in the present work are in a good agreement with the literature data [9], the linear compressibility values significantly differ from those reported previously [9] (Table I).

The pressure dependences of $B_{13}N_2$ and $B_{50}N_2$ unit-cell volumes to 30 GPa are plotted in Fig. 4. A least-squares fit using the Murnaghan equation of state [22]:

$$P(V) = \frac{B_0}{B_0'}\left[\left(\frac{V}{V_0}\right)^{B_0'} - 1\right] \tag{3}$$

($B_0$ and $B_0'$ are bulk modulus and its first pressure derivative, respectively) give the bulk modulus values (Table I). Bulk modulus of $B_{50}N_2$ (167 GPa) is significantly smaller than that of $B_{13}N_2$ (205 GPa). Such difference in compressibilities of two boron subnitrides can be explained by the fact that their structures derives from two different boron allotropes i.e. α-tetragonal boron (t′-$B_{52}$) and α-rhombohedral boron (α-$B_{12}$), respectively. According to *ab initio* calculations, α-tetragonal boron is significantly more compressible ($B_0$ = 162 GPa [23]) than α-$B_{12}$ ($B_0$ = 213 GPa [24]). In accordance with this, $B_{50}N_2$ is less dense and more compressible phase.

The bulk moduli of boron subnitrides, boron allotropes (α-$B_{12}$ [24,25], β-$B_{106}$ [24,26,27], γ-$B_{28}$ [28,29], t′-$B_{52}$ [23]) and some boron-rich compounds with structures related to α-rhombohedral boron [9,30-36] are summarized in Table IV and Fig. 5. Only data from *in situ* X-ray diffraction experiments in DACs are presented, with the exception made for $B_{12}As_2$: the experimental $B_0$-value of 216 GPa [36] is obviously overestimated due to the non-hydrostatic conditions in a DAC above 8 GPa (methanol-ethanol pressure transmitting medium), so we considered the theoretically predicted value of $B_{12}As_2$ bulk modulus i.e. 182 GPa [37].

One can see that bulk modulus of $B_{13}N_2$ is in good agreement with the previously reported value [9] and is of the same order of magnitude as those for α-$B_{12}$, $B_{12}O_2$, $B_{12}P_2$ and $B_{12}As_2$, while tetragonal boron subnitride $B_{50}N_2$ is obviously the most compressible phase among all considered boron-rich compounds.

Finally, no evidence for room-temperature transformation(s) into another crystalline structure or an amorphous phase has been observed for both boron subnitrides over the whole pressure range under study.

## IV. Conclusions

New tetragonal boron subnitride, $B_{50}N_2$, has been studied by powder X-ray diffraction at ambient conditions and at pressures to 30 GPa. The Rietveld refinement of the powder diffraction data have shown that crystal structure of $B_{50}N_2$ belongs to the $P\bar{4}n2$ space group with lattice parameters $a_0$ = 8.8181(2) Å and $c_0$ = 5.0427(10) Å. The compressibilities of two boron subnitrides, $B_{50}N_2$ and $B_{13}N_2$, have been measured at room temperature in the same experiment. No pressure-induced phase transitions have been observed up to 30 GPa. The bulk moduli of $B_{50}N_2$ and $B_{13}N_2$ have been determined to be 167(2) and 205(2) GPa, respectively, with first bulk modulus pressure derivatives fixed to 4.0.

**Acknowledgements**

The authors thank Drs. Yann Le Godec and Alain Polian (IMPMC) for help in DAC preparation, Drs. Paolo Lotti and Boby Joseph (Elettra) for assistance in synchrotron experiments, and Dr. Thierry Chauveau (LSPM) for aid in Rietveld analysis. Synchrotron X-ray diffraction studies were carried out during beam time allocated to Proposal 20160061 at Elettra Sincrotrone Trieste. This work was financially supported by European Union's Horizon 2020 Research and Innovation Programme under Flintstone2020 project (grant agreement No 689279).

Table I. Lattice parameters, linear compressibilities, axial and bulk moduli and their first pressure derivatives of boron subnitrides.

| | $B_{50}N_2$ | $B_{13}N_2$ | $B_{13}N_2$ [9] |
|---|---|---|---|
| Space group | $P\bar{4}n2$ | $R\bar{3}m$ | $R\bar{3}m$ |
| $a_0$, Å | 8.8181(2) | 5.4537(3) | 5.4585 |
| $c_0$, Å | 5.0427(10) | 12.2537(7) | 12.2530 |
| $V_0$, Å$^3$ | 392.11(10) | 315.62(6) | 316.16 |
| ρ, g/cm$^3$ | 2.408(1) | 2.660(1) | 2.656 |
| $k_a \times 10^{-3}$, GPa$^{-1}$ | 1.67(11) | 1.43(6) | 1.56(2) |
| $k_c \times 10^{-3}$, GPa$^{-1}$ | 2.40(9) | 2.41(9) | 1.80(4) |
| $\beta_{0,a}$, GPa | 600(39) | 700(28) | 640(10) |
| $\beta_{0,c}$, GPa | 416(15) | 415(15) | 558(8) |
| $B_0$, GPa | 167(2) | 205(2) | 200(15) |
| $B_0'$ | 4.0$^f$ | 4.0$^f$ | 4.0$^f$ |

$^f$ fixed value

Table II Lattice parameters and unit cell volumes of boron subnitrides *versus* pressure at room temperature.

| Pressure, GPa | *a*, Å | *c*, Å | V, $Å^3$ |
|---|---|---|---|
| $B_{50}N_2$ | | | |
| 0.0 | 8.8181 | 5.0427 | 392.11 |
| 7.2 | 8.6995 | 4.9630 | 375.61 |
| 10.9 | 8.6804 | 4.9253 | 371.12 |
| 13.7 | 8.6316 | 4.9001 | 365.08 |
| 17.4 | 8.6000 | 4.8779 | 360.77 |
| 19.6 | 8.5633 | 4.8545 | 355.98 |
| 22.3 | 8.5400 | 4.8368 | 352.76 |
| 25.3 | 8.5150 | 4.8155 | 349.15 |
| 27.5 | 8.4713 | 4.8034 | 344.71 |
| 29.4 | 8.4608 | 4.7800 | 342.18 |
| $B_{13}N_2$ | | | |
| 0.0 | 5.4537 | 12.2537 | 315.62 |
| 7.2 | 5.4000 | 12.0805 | 305.06 |
| 10.9 | 5.3765 | 12.0000 | 300.40 |
| 13.7 | 5.3545 | 11.9500 | 296.70 |
| 17.4 | 5.3379 | 11.9100 | 293.88 |
| 19.6 | 5.3230 | 11.8731 | 291.34 |
| 22.3 | 5.3013 | 11.8269 | 287.84 |
| 25.3 | 5.2847 | 11.7911 | 285.18 |
| 27.5 | 5.2724 | 11.7788 | 283.55 |
| 29.4 | 5.2700 | 11.7475 | 282.54 |

Table III. The atomic structure data for tetragonal $B_{50}N_2$.

| Atom label | Wyckoff position | *x* | *y* | *z* | Site occupancy |
|---|---|---|---|---|---|
| N1 | 2b | 0.0000 | 0.0000 | 0.4996(5) | 0.995(13) |
| B1 | 2a | 0.0000 | 0.0000 | 0.0000 | 0.922(20) |
| B2 | 8i | 0.3280 | 0.0950 | 0.3950 | 1.0[f] |
| B3 | 8i | 0.0950 | 0.3280 | 0.3950 | 1.0[f] |
| B4 | 8i | 0.2230 | 0.0780 | 0.1050 | 1.0[f] |
| B5 | 8i | 0.0780 | 0.2230 | 0.1050 | 1.0[f] |
| B6 | 8i | 0.1270 | 0.1270 | 0.3950 | 1.0[f] |
| B7 | 8i | 0.2500 | 0.2500 | -0.0780 | 1.0[f] |

[f] the atom site occupancies were fixed to 1.0

Table IV. Bulk moduli and their first pressure derivatives of boron allotropes and boron-rich solids with structures related to $\alpha$-rhombohedral boron

| Compound | $B_0$, GPa | $B_0'$ | Ref. |
|---|---|---|---|
| $\alpha$-$B_{12}$ | 213(15) | 4.0$^f$ | [24] |
| | 224(7) | 3.0(3) | [25] |
| $\beta$-$B_{106}$ | 185(7) | – | [24] |
| | 205(16) | 4.3(1.6) | [26] |
| | 210(6) | 2.2 | [27] |
| $\gamma$-$B_{28}$ | 237(5) | 2.7(3) | [28] |
| | 227(3) | 2.5(2) | [29] |
| $B_{12}C_3$ | 199(7) | 1(2) | [30] |
| | 243(3) | 3.6(2) | [31] |
| | 222(2) | 4.0$^f$ | [32] |
| | 239(7) | 3.2(3) | |
| $B_{13}N_2$ | 200(15) | 4.0$^f$ | [9] |
| | 205(2) | 4.0$^f$ | present study |
| $B_{50}N_2$ | 167(3) | 4.0$^f$ | present study |
| $B_{12}O_2$ | 181(5) | 6.0 | [33] |
| $B_{12}P_2$ | 192(11) | 5.5(12) | [34] |
| | 207(7) | 6.6(8) | [35] |
| $B_{12}As_2$ | 216(3) | 2.2(3) | [36] |
| | 182* | – | [37] |

$^f$ fixed value

* theoretically predicted value

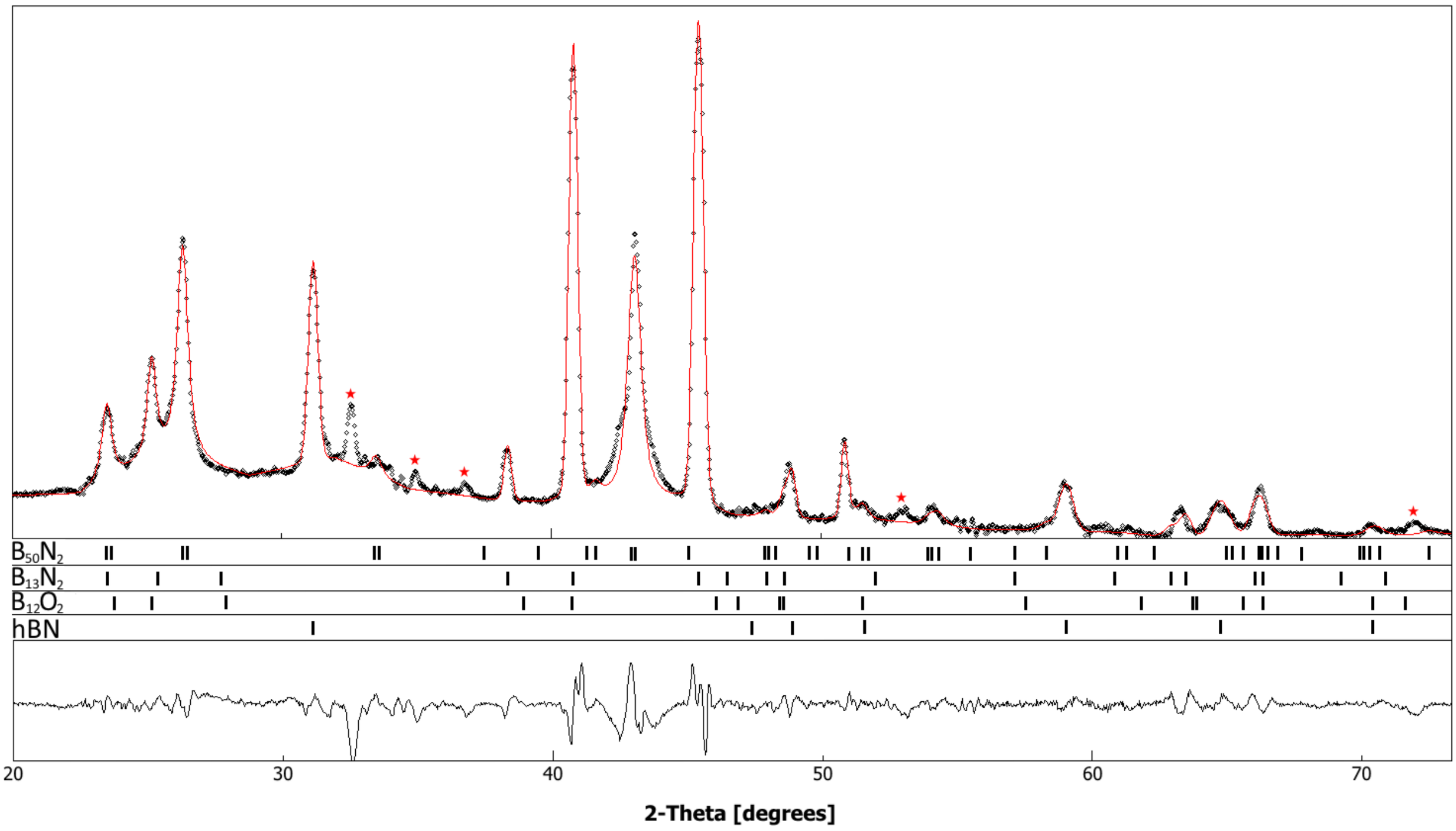

Fig. 1. Rietveld full profile refinement of powder X-ray diffraction pattern of the recovered sample (diffraction lines of impurities are marked by red stars).

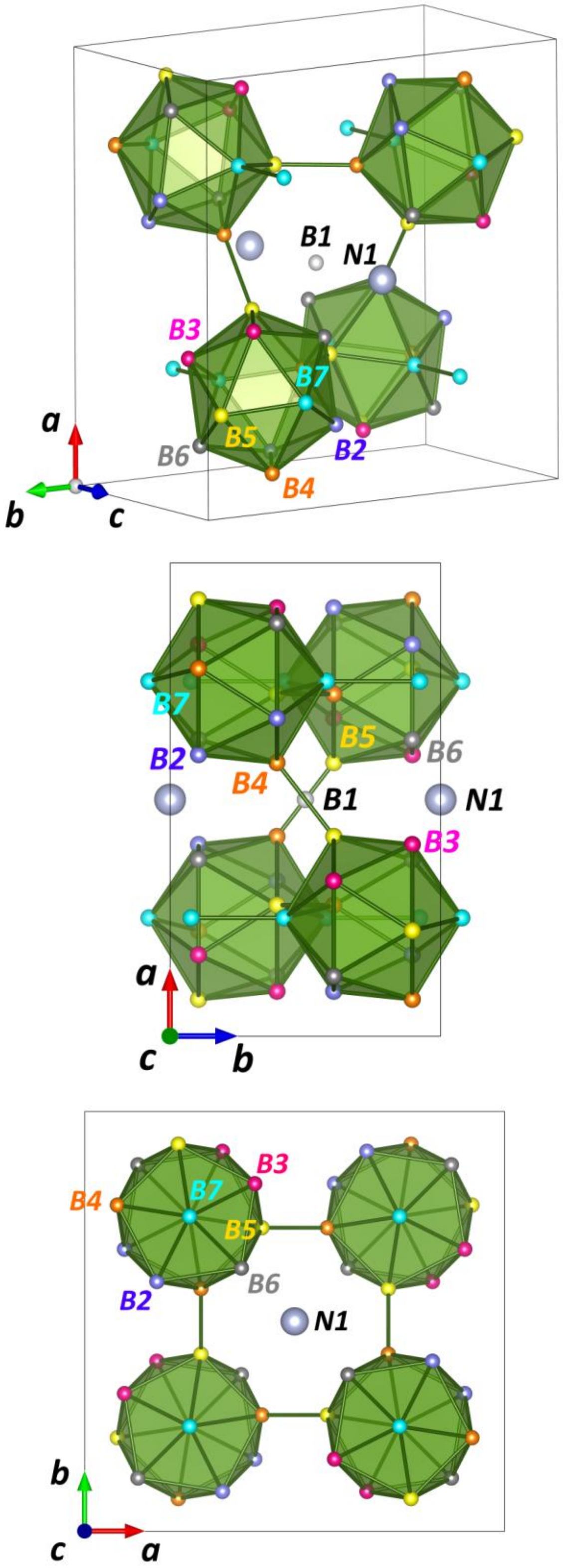

Fig. 2. Crystal structure of new tetragonal boron subnitride $B_{50}N_2$ (B and N atoms in crystallographically independent positions are marked in different colors).

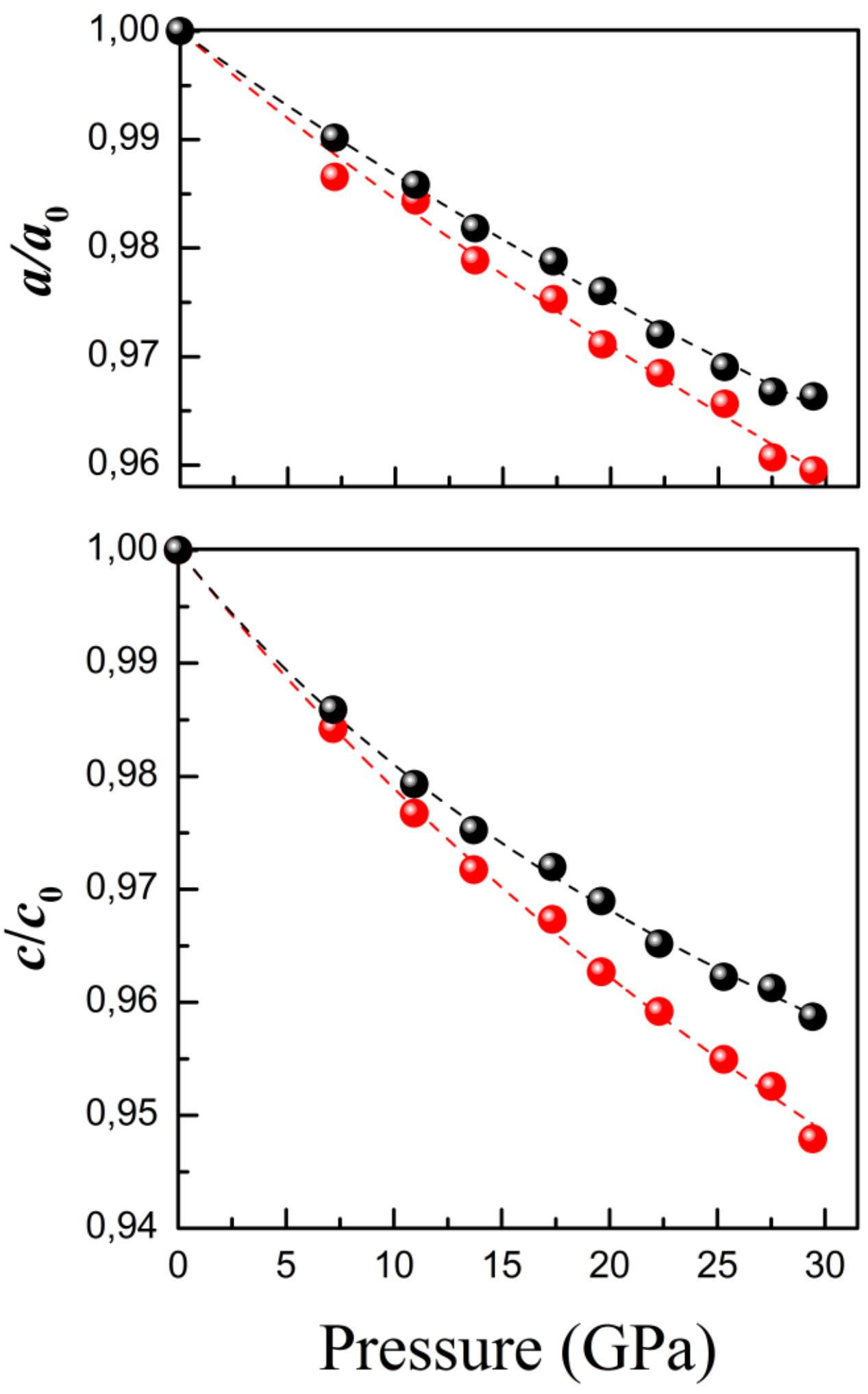

Fig. 3. Relative lattice parameters of $B_{50}N_2$ (red circles) and $B_{13}N_2$ (black circles) *versus* pressure. The dashed lines represent the corresponding fits of one-dimensional analog of the Murnaghan equation of state to the experimental data.

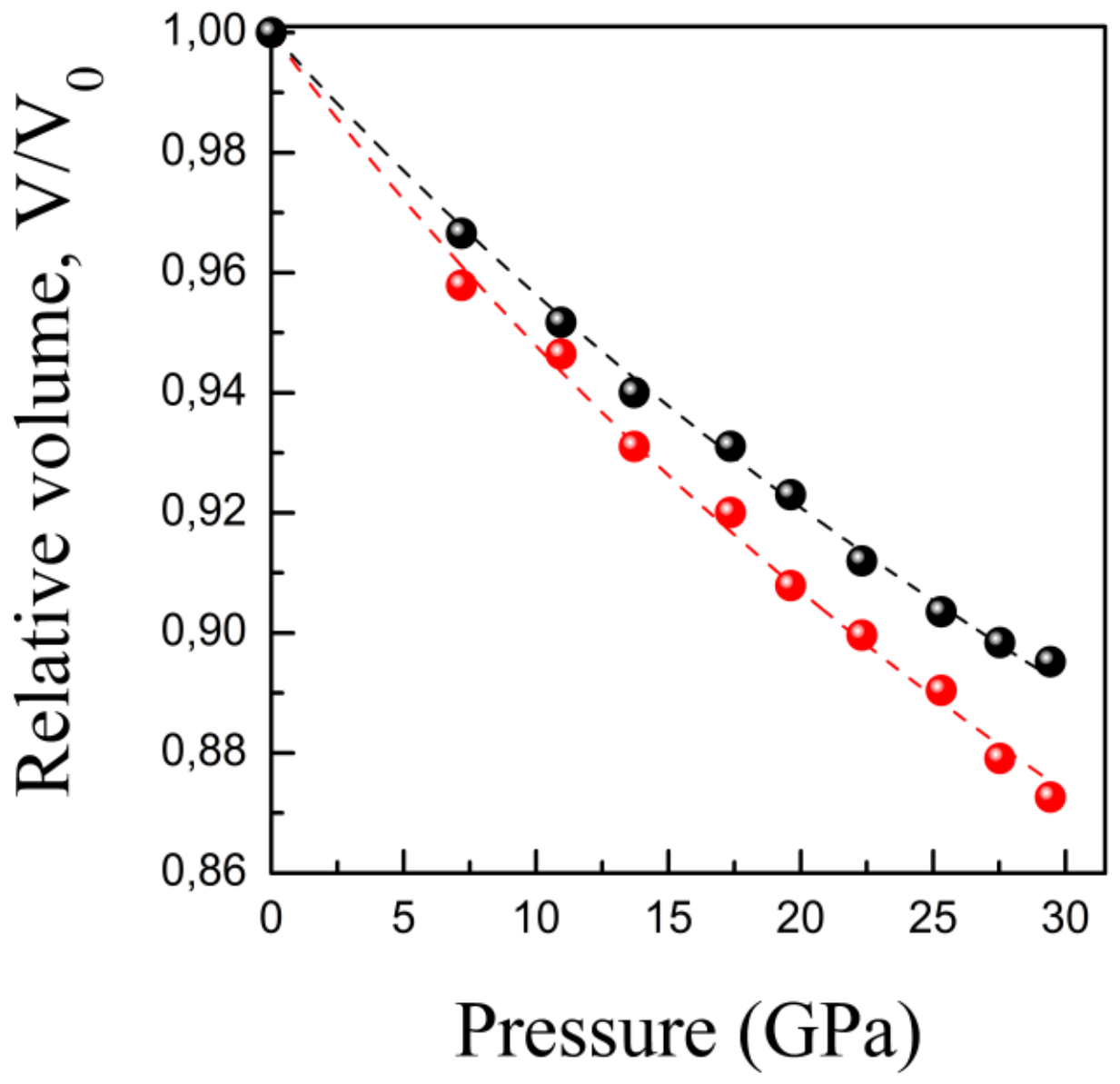

Fig. 4. Equations of state of $B_{13}N_2$ (black circles) and $B_{50}N_2$ (red circles). The dashed lines represent the Murnaghan fits to the experimental data.

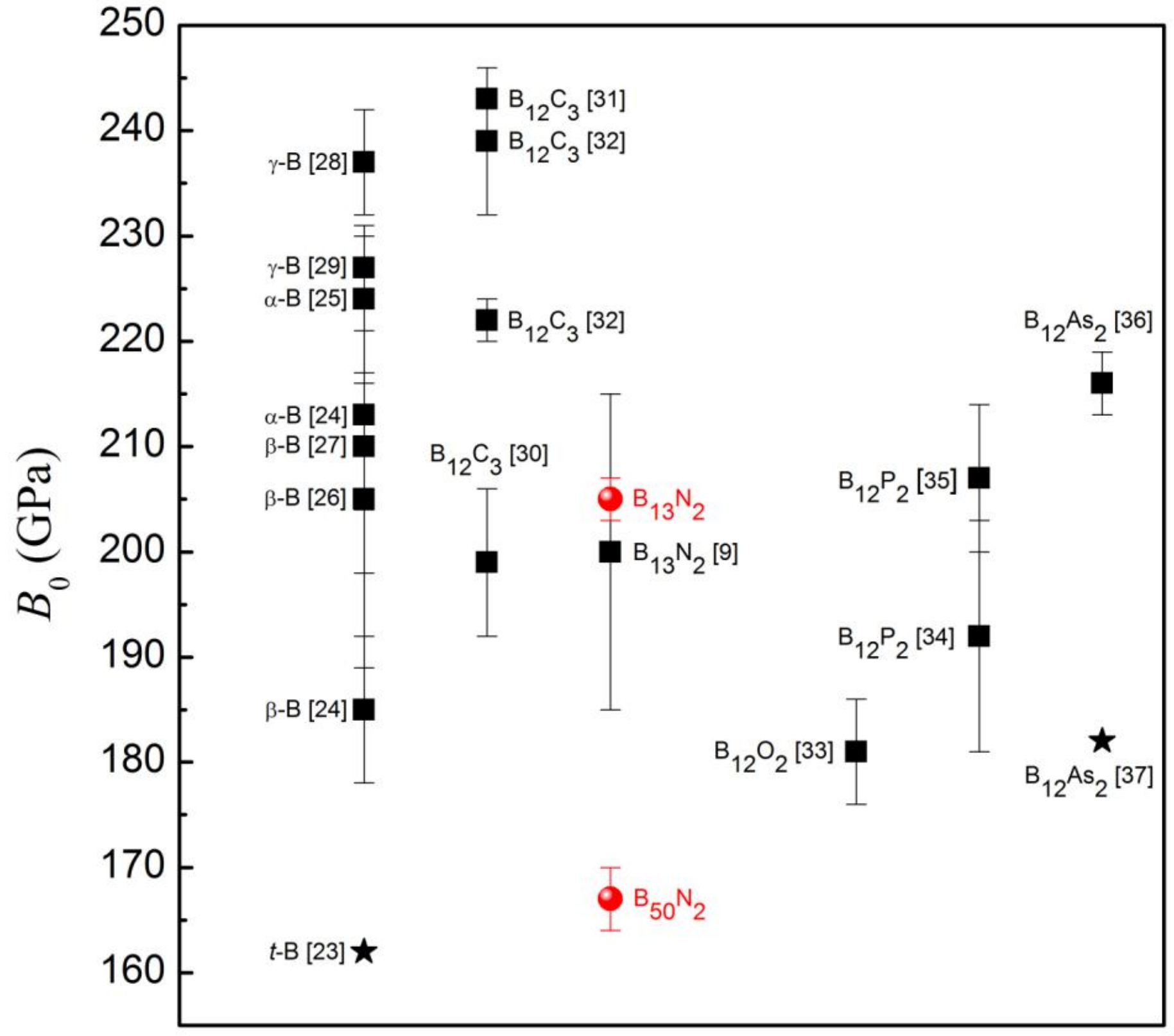

Fig. 5. Bulk moduli of boron allotropes and boron-rich compounds: black squares and red circles represent literature and our experimental data, respectively. The theoretical predictions of α-tetragonal boron and $B_{12}As_2$ bulk moduli are shown by black stars.